\documentclass[12pt]{article}
\usepackage{amsfonts,latexsym,epsfig,amscd}
\usepackage{amssymb}

\newcommand{\R}{{\mathbb{R}}}

\newcommand{\I}{{\mathbb{I}}}

\newcommand{\be}{\begin{equation}}
\newcommand{\ee}{\end{equation}}
\newcommand{\bea}{\begin{eqnarray}}
\newcommand{\eea}{\end{eqnarray}}
\newcommand{\bean}{\begin{eqnarray*}}

\newcommand{\eean}{\end{eqnarray*}}
\font\upright=cmu10 scaled\magstep1

\newcommand{\PP}{\hbox{\upright\rlap{I}\kern 1.5pt P}}

\newcommand{\identity}{{\upright\rlap{1}\kern 2.0pt 1}}

\newcommand{\HH}{\mbox{\hbox{\upright\rlap{I}\kern 1.7pt H}}}

\newcommand{\fr}{\frac}
\newcommand{\lm}{\lambda}
\newcommand{\ra}{\rightarrow}
\newcommand{\al}{\alpha}

\newcommand{\bt}{\beta}
\newcommand{\pr}{\partial}

\newcommand{\hs}{\hspace{5mm}}

\newcommand{\dg}{\dagger}

\newcommand{\acc}{\\[3mm]}

\begin{document}
\begin{titlepage}
\strut\hfill
\vspace{0mm}
\begin{center}

{\Large \bf Skyrmions, Rational Maps \& Scaling Identities}
\vspace{12mm}

{\bf E. G. Charalampidis${}^{1^{*}}$,  \ T. A. Ioannidou${}^{1^{\dg}}$ 
\ and\ N. S. Manton${}^{2^{\ddagger }}$}
\\[12mm]
$^1${\footnotesize Department of Mathematics, Physics and
Computational Sciences, Faculty of Engineering,\\\vspace{-2mm}
Aristotle University of Thessaloniki, GR-54124 Thessaloniki, Greece }\acc
$^2${\footnotesize Department of Applied Mathematics and Theoretical 
Physics, University of Cambridge, \\\vspace{-2mm}
Wilberforce Road, Cambridge CB3 0WA, UK}
\acc

\vspace{12mm}

\begin{abstract}
Starting from approximate Skyrmion solutions obtained using the
rational map ansatz, improved approximate Skyrmions are constructed 
using scaling arguments. Although the energy improvement is small, 
the change of shape clarifies whether the true Skyrmions are more 
oblate or prolate.
\end{abstract}
\end{center}

\vspace{50mm}

$^*${\footnotesize Email: {\tt echarala@auth.gr}} \vspace{-2mm}

$^\dg${\footnotesize Email: {\tt ti3@auth.gr}} \vspace{-2mm}

$^\ddagger${\footnotesize Email: {\tt N.S.Manton@damtp.cam.ac.uk}}

\end{titlepage}

\section{Skyrmions and Rational Maps}

The Skyrme model \cite{Sk} is an effective field theory used to describe 
nuclei. For static fields, the classical solutions of the Skyrme model are
stationary points (either minima or saddle points) of the energy functional
\be
E=\frac{1}{12\pi^2}\int_{\R^3}\left\{-\fr{1}{2}\mbox{tr}\left(R_i^2\right)
-\fr{1}{16}\mbox{tr}\left(\left[R_i,\;R_j\right]^2\right)
+m^2\,\mbox{tr}(\I-U)\right\}d^3{\bf x} \,,
\label{gene}
\ee
where ${\bf x}=(x_1,x_2,x_3)$ are the Cartesian coordinates, 
and $U({\bf x})$ is an $SU(2)$-valued scalar field, known as the Skyrme 
field, which describes pions nonlinearly. The gradient of the Skyrme
field is captured by the current components $R_i=\pr_iU\, U^{-1} \ 
(i=1,2,3)$. In (\ref{gene}), the energy and length units have been 
scaled away, leaving only the pion mass parameter $m$, which is the 
physical pion mass in scaled units. [In many past studies the pion
mass term $m^2 \mbox{tr}(\I-U)$ was omitted.] In this study, when we
consider massive pions, we assume that $m=1$, since only for $m$ at or close
to this value can the proton and delta masses be reproduced (for
more details, see \cite{BSm1} and References therein). Also, when
$m=1$, nuclei are represented by Skyrmions that have a fairly uniform
density, rather than a hollow, roughly spherical structure \cite{BSm2,BMS}.  

Finiteness of the energy requires that $U$ 
approaches the identity matrix $\I$ at spatial infinity (and this boundary
condition is also imposed if $m=0$). Hence, $\R^3$ is
topologically compactified to a large 3-sphere. $U$ is then a mapping from 
$S^3\mapsto SU(2)$, and is classified by the integer-valued
degree (winding number)
\be
B=-\fr{1}{4\pi^2}\int_{\R^3} \mbox{tr}\left(R_1 R_2 R_3\right)
d^3{\bf x} \,,\label{bar}
\ee
which is a topological invariant. 
In particular, $B$ classifies the solitonic sectors of the model 
and in applications to nuclei, $B$ is identified
with the  baryon number of the Skyrme field configuration. A Skyrmion
is a static soliton of minimal energy for given $B$, although 
sometimes, more loosely, a local minimum or saddle point of the 
energy functional is also called a Skyrmion.

In \cite{HMS}, minimal energy Skyrmions with
massless pions were approximated by an ansatz involving a rational 
map $R$ between Riemann spheres. Here, a point in $\R^3$ 
is labelled by coordinates $(r,z)$ where $r$ is the radial 
distance from the origin and $z=\tan(\theta/2)\,e^{i\varphi}$ 
specifies the direction from the origin ($\theta,\varphi$ are the
usual spherical polar coordinates). Let $R$ be a degree $N$ 
rational map of the form $R(z)=\fr{p\,(z)}{q\,(z)}$, where $p$ 
and $q$ are polynomials in $z$ with no common factors, such that 
$\mbox{max}\left[\deg(p),\deg(q)\right]=N$.
Then the ansatz for the Skyrme field is 
\be
U(r,z)=\mbox{exp}\left[\fr{if(r)}{1+|R(z)|^2}
\left(\begin{array}{ll}
1-|R(z)|^2&\hs 2\overline{R(z)}\\\hs2R(z)&|R(z)|^2-1
\end{array}\right)\right] \,,\label{rm}
\ee
where $f(r)$ is a real profile function satisfying the boundary 
conditions $f(0)=k\pi$ (for some integer $k$) and $f(\infty)=0$.
It is straightforward to verify that the baryon number of this field 
is $B=kN$. In the remainder of this paper we only consider $k=1$, thus $B=N$.
An attractive feature of the rational map ansatz (\ref{rm}) is 
that it leads to a simple energy expression which is minimized 
by first minimizing with respect to the parameters in the rational 
map $R(z)$, and then determining the profile 
function $f(r)$ by solving an ordinary differential
equation. This approach gives close approximations to true 
Skyrmions. Clearly, this procedure encounters difficulties when there 
are two or more Skyrmion solutions, either saddle points or genuine minima,
with very similar energies. The ansatz might lead to the wrong energy ordering.

For massless pions the minimal energy Skyrmions up to $B = 22$ are all 
well-approximated by the rational map ansatz \cite{BSf}. However, the 
inclusion of the pion mass term in the energy functional makes the 
Skyrme model more realistic in describing real nuclei. The solutions 
change dramatically when $m=1$, especially for large baryon number 
($B\geq 8$). Some Skyrmions are known to be prolate (cigarlike) 
or oblate (pancakelike), for example, those with baryon numbers $B=8$ 
and $B=12$, respectively \cite{BMS}. Spherical shell-like configurations 
given by the rational map ansatz are still useful starting points 
in the search for true solutions, but they are unstable \cite{BSm2}, 
because their hollow interiors have substantial potential energy when 
$m=1$, and the fields relax to more compact structures, some being 
rather flat. This way the Skyrmion reduces its volume to surface area
ratio and thus reduces its energy. Moreover, the Skyrmions with
massive pions are smaller in size and exponentially localized, 
compared to Skyrmions with massless pions which are algebraically 
localized. This can be seen in Figure \ref{fig:Bdensities}.

\section{Independent Length Rescalings}

Recently, Manton \cite{Manton} studied the effect of independent length
rescalings in the three Cartesian directions in the Skyrme model. 
The simple Derrick scaling identity \cite{Der} and
further novel identities were derived, relating 
contributions to the total energy of a Skyrmion. For an exact 
Skyrmion solution, the identities are satisfied exactly.  
 
In what follows we will study the effect of scaling manipulations on the energy
functional for various Skyrme field configurations which are not exact
solutions, for both massless and massive pions. In particular, for an
approximate Skyrmion described by the rational map ansatz, we can
lower the energy by a rescaling of the form
\be
 x_1\ra \lm_1 x_1 \,,\hs x_2\ra \lm_2 x_2 \,,\hs x_3\ra \lm_3 x_3 \,,
\ee
with the parameters $\lm_i$ not all the same. 
For a general Skyrme field, $U(x_1,x_2,x_3)$ is replaced under such a
rescaling by 
$\widetilde{U}(x_1,x_2,x_3)=U(\lm_1 x_1,\lm_2 x_2,\lm_3 x_3)$ and the energy
$\widetilde{E}$ of the rescaled field $\widetilde{U}({\bf x})$ is the
modified version of the original energy (\ref{gene}),
 \begin{eqnarray}
 \widetilde{E}  \!\!\!\!& = & \!\!\!\!\frac{1}{12\pi^{2}}\!\!\int_{\R^3}\!\! {\Bigg\lbrace\!\!\! -\frac{1}{2}\frac{\lambda_{1}}{\lambda_{2}\lambda_{3}}\mbox{tr}\left(R_{1}^{2}\right) -\frac{1}{2}\frac{\lambda_{2}}{\lambda_{3}\lambda_{1}}\mbox{tr}\left(R_{2}^{2}\right) -\frac{1}{2}\frac{\lambda_{3}}{\lambda_{1}\lambda_{2}}\mbox{tr}\left(R_{3}^{2}\right)-\frac{1}{8}\frac{\lambda_{1}\lambda_{2}}{\lambda_{3}}\mbox{tr}\left(\left[R_{1},R_{2}\right]^{2}\right)} \nonumber \\
    \!\!\!&  &
    \hs\hs\ \  \,-\frac{1}{8}\frac{\lambda_{2}\lambda_{3}}{\lambda_{1}}\mbox{tr}\left(\left[R_{2},R_{3}\right]^{2}\right)-\frac{1}{8}\frac{\lambda_{3}\lambda_{1}}{\lambda_{2}}\mbox{tr}\left(\left[R_{3},R_{1}\right]^{2}\right)+\frac{m^2}{\lambda_{1}\lambda_{2}\lambda_{3}}\mbox{tr}\left(\I-U\right)\Bigg\rbrace d^{3}{\bf x} \,,
\label{rene}
\end{eqnarray}
where the current components $R_i$ are evaluated for the original 
field $U({\bf x})$. 

For an exact Skyrmion solution $U({\bf x})$,
$\widetilde{E}$ is stationary with respect to $\lm_i$ at $\lm_i=1$
(since the energy is stationary with respect to any smooth change of the
field that preserves the boundary condition).
As in \cite{Manton}, we may restrict to special rescalings which 
together span all possibilities. In particular, a rescaling in the 
$(x_2,x_3)$ plane with no rescaling of $x_1$ can be achieved by 
setting $\lm_2=\lm_3=\lm$ and $\lm_1=1$ in (\ref{rene}). 
Then the derivative of $\widetilde{E}$ with respect to $\lm$ vanishes at
$\lm=1$. This gives identity (\ref{I1}). Similarly, by permutation, 
the following identities are obtained:
\bea
{\cal I}_1:
&&\int_{\R^3}\left\{-\fr{1}{2}\,\mbox{tr}\left(R_1^2\right)
+\fr{1}{8}\,\mbox{tr}\left([R_2,R_3]^2\right)
+m^2\mbox{tr}\left(\I-U\right)\right\}d^3{\bf
  x}=0\,, \label{I1}\\
{\cal I}_2:
&&\int_{\R^3}\left\{-\fr{1}{2}\,\mbox{tr}\left(R_2^2\right)
+\fr{1}{8}\,\mbox{tr}\left([R_3,R_1]^2\right)
+m^2\mbox{tr}\left(\I-U\right)\right\}d^3{\bf x}=0\,,\label{I2}\\
{\cal I}_3:
&&\int_{\R^3}\left\{-\fr{1}{2}\,\mbox{tr}\left(R_3^2\right)
+\fr{1}{8}\,\mbox{tr}\left([R_1,R_2]^2\right)
+m^2\mbox{tr}\left(\I-U\right)\right\}d^3{\bf x}=0 \,.\label{I3}
\eea
We refer to the left hand sides of these identities as the scaling
integrals $\{{\cal I}_i : i=1,2,3 \}$.
The sum of the above three identities is Derrick's identity for
Skyrmions, which is also obtained by considering the uniform 
rescaling $\lm_1=\lm_2=\lm_3=\lm$.
 
In \cite{BSf}, it has been shown that for $m=0$ and $B\geq 7$, there are
increasingly many Skyrmions, with different symmetries and
different shapes, whose energies are very close to the minimal value. The 
additional configurations are local minima or saddle points of the 
Skyrme energy, whose energies are difficult to distinguish
numerically. In particular for $B=9$ there are two Skyrmion
possibilities with $D_{4d}$ and $T_d$ symmetries; the first is
probably the global minimum and the second a saddle point. For $B=10$ 
there are at least four solutions close to minimal energy, but for 
$B=11$ the minimal energy Skyrmion appears isolated. It can be difficult 
to ensure numerically that the fields have fully relaxed to a
solution. One good way to test if they have would be to check whether 
the scaling identities (\ref{I1})-(\ref{I3}) are satisfied.

The rational map ansatz gives good {\it approximations} to
exact Skyrmion solutions, which generally do not satisfy all these scaling
identities. An optimal rescaling should deform the rational map 
configurations closer to the exact Skyrmions.
Moreover, after rescaling, the scaling identities will all be satisfied. 
The argument is as follows. We start with the approximate solution and
evaluate the seven contributions to the energy that occur in (\ref{gene}), and
with modified coefficients in (\ref{rene}). We then find the parameters
$\lm_i$ that minimize (\ref{rene}). This tells us how to rescale
the initial field so as to obtain an improved approximate
solution. The rescaled field satisfies the identities
(\ref{I1})-(\ref{I3}), as it has minimal energy with respect to
further rescalings. An alternative formulation of this, more
convenient computationally, is to say that after rescaling, 
the identities 
\bea
\widetilde{\cal I}_1:
&&\int_{\R^3}\left\{-\fr{1}{2}\,
\frac{\lambda_{1}}{\lambda_{2}\lambda_{3}}\mbox{tr}\left(R_1^2\right)
+\fr{1}{8}\,\frac{\lambda_{2}\lambda_{3}}{\lambda_{1}}
\mbox{tr}\left([R_2,R_3]^2\right)
+\frac{m^2}{\lambda_{1}\lambda_{2}\lambda_{3}}
\mbox{tr}\left(\I-U\right)\right\}d^3{\bf x}=0\, , \label{tildeI1}\\
\widetilde{\cal I}_2:
&&\int_{\R^3}\left\{-\fr{1}{2}\,
\frac{\lambda_{2}}{\lambda_{3}\lambda_{1}}\mbox{tr}\left(R_2^2\right)
+\fr{1}{8}\,\frac{\lambda_{3}\lambda_{1}}{\lambda_{2}}
\mbox{tr}\left([R_3,R_1]^2\right)
+\frac{m^2}{\lambda_{1}\lambda_{2}\lambda_{3}}
\mbox{tr}\left(\I-U\right)\right\}d^3{\bf x}=0\, ,\label{tildeI2}\\
\widetilde{\cal I}_3:
&&\int_{\R^3}\left\{-\fr{1}{2}\,
\frac{\lambda_{3}}{\lambda_{1}\lambda_{2}}\mbox{tr}\left(R_3^2\right)
+\fr{1}{8}\,\frac{\lambda_{1}\lambda_{2}}{\lambda_{3}}
\mbox{tr}\left([R_1,R_2]^2\right)
+\frac{m^2}{\lambda_{1}\lambda_{2}\lambda_{3}}
\mbox{tr}\left(\I-U\right)\right\}d^3{\bf x}=0 \label{tildeI3}
\eea
are satisfied, where $\lambda_{1},\lambda_{2},\lambda_{3}$ are the
parameters we find, and $U$, $R_1$, $R_2$ and $R_3$ are the original 
field and current components before rescaling.

The energy is lowered by this rescaling, so one gets closer 
to an exact solution. More interesting, 
perhaps, than the small reduction of energy is the small change 
in shape. We learn from the calculation whether the true Skyrmion is 
more prolate or oblate than the approximate Skyrmion (which itself 
is rather round), or triaxial. 

Note that in all cases, the product of the rescaling parameters is 
very close to unity, i.e. $\lm_1\lm_2\lm_3=1$, if the starting 
point is the optimised rational map ansatz with the profile function 
worked out numerically. This is because the starting point satisfies 
Derrick's identity.

After rescaling a Skyrme field, the baryon density becomes
\begin{equation}
\widetilde{\cal B}({\bf x})=-\frac{\lm_1\lm_2\lm_3}{4\pi^{2}}\,
\mbox{tr}\left(R_{1}R_{2}R_{3}\right) \,,
\label{rbar}
\end{equation}
where the right hand side denotes $\lm_1\lm_2\lm_3$ times the original
baryon density evaluated at $\widetilde{\bf x} = 
(\lm_1 x_1,\lm_2 x_2,\lm_3 x_3)$. Its integral is of course
unchanged. In particular, if $\lm_1\lm_2\lm_3=1$, the baryon density 
transforms as a scalar quantity, so that the new density at ${\bf x}$ 
is the old density at $\widetilde{\bf x}$.

In order to make some practical use of the scaling identities
(\ref{I1})-(\ref{I3}) it would be best to work with Skyrmions 
that are known to be, or are expected to be, very far from spherically 
symmetric. Examples are the Skyrmions with $m=1$ and baryon number 
a multiple of four, composed of $B=4$ subunits \cite{BMS} as in 
the $\al$-particle model of nuclei. These Skyrmions sometimes look 
like part of the infinite Skyrme crystal.

We will calculate here the optimal rescaling of approximate Skyrmions
given by the rational map ansatz, for $B=6,9,10$ and $11$, and 
for both $m=0$ and $m=1$. In some of these cases the exact solutions are
known, so we do not learn much new. For $B=9$ and $B=11$ the exact 
solutions for $m=1$ are not known, so here we get some insight into 
the shapes of the true Skyrmions. Since the rescalings are all quite 
close to unity (within a few percent), it is important that the 
calculations of the parameters $\lm_i$ are not dominated by numerical 
errors in the energy integrals. In fact, the numerical errors in the 
rational map approach are smaller than $1\%$.

\begin{figure}[pht]
  \centering  
  \includegraphics[width=4cm]{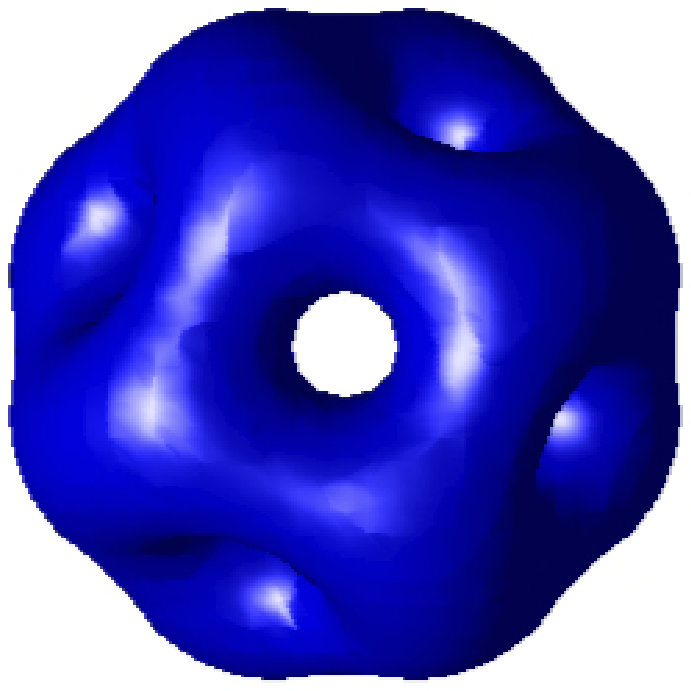}
    \includegraphics[width=4cm]{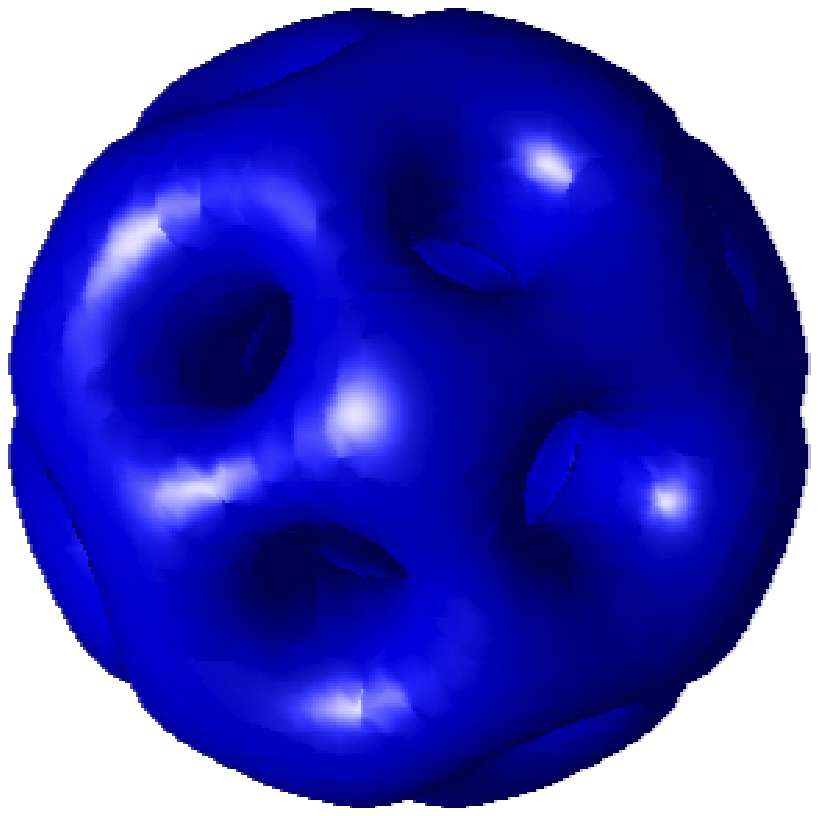}
      \includegraphics[width=4cm]{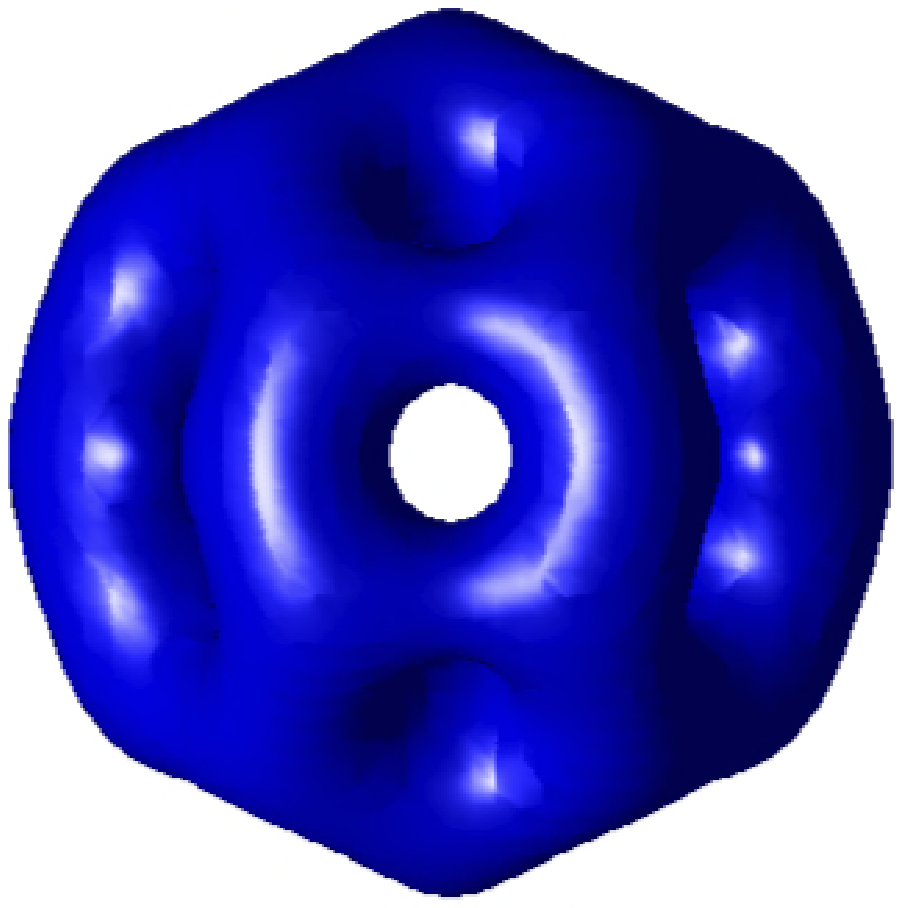}
       \includegraphics[width=4cm]{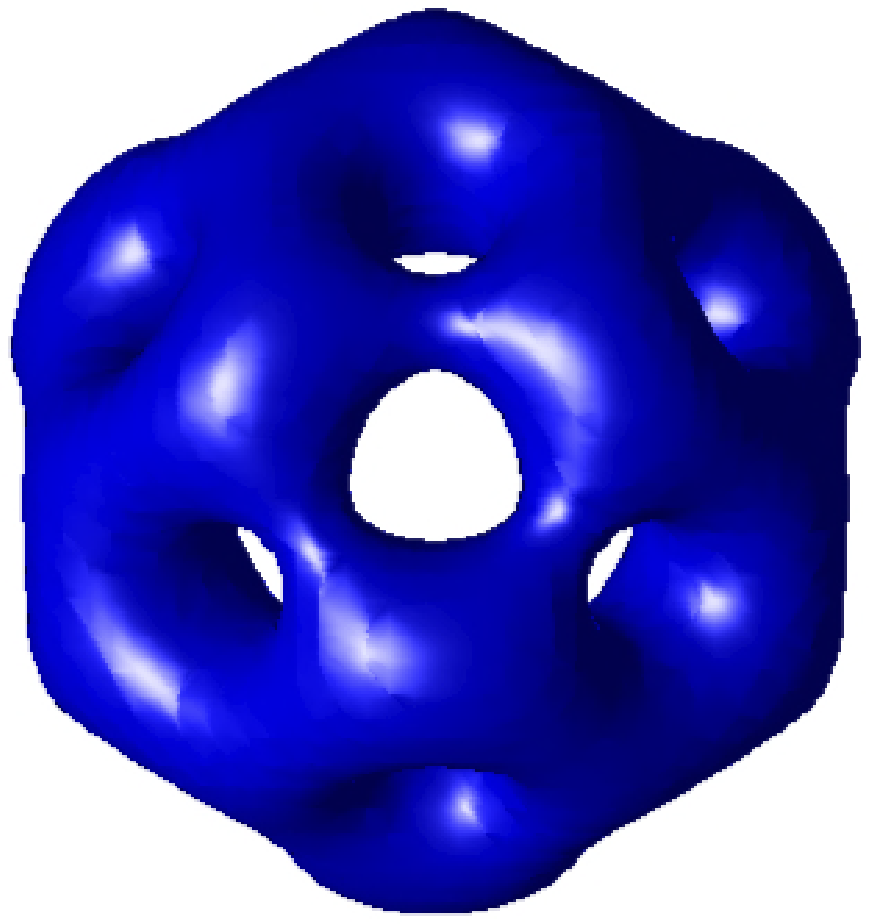}\acc

 \includegraphics[width=4cm]{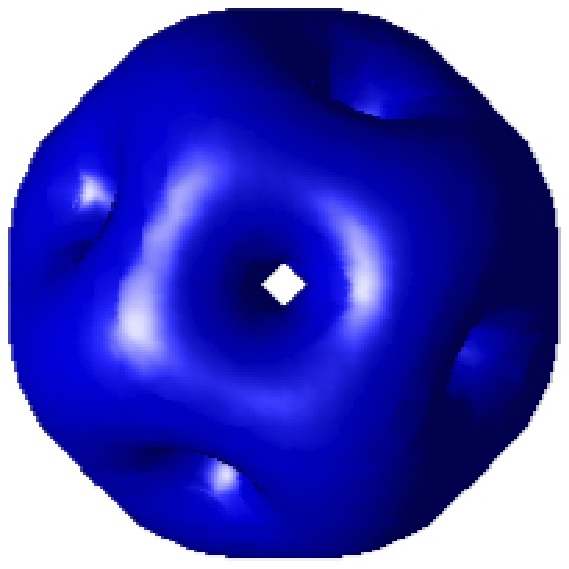}
  \includegraphics[width=4cm]{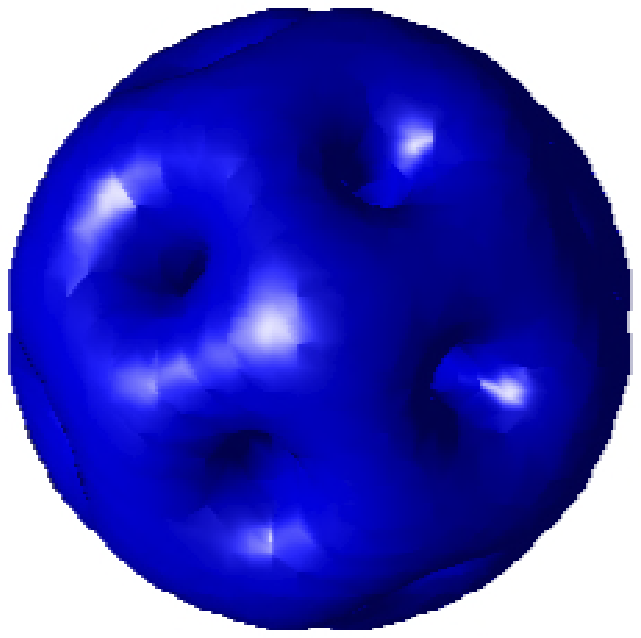}
   \includegraphics[width=4cm]{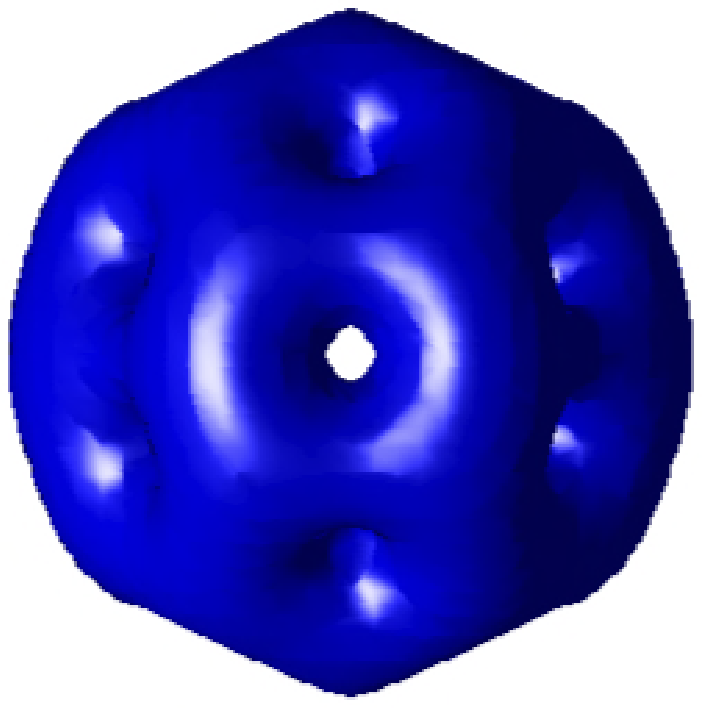}
    \includegraphics[width=4cm]{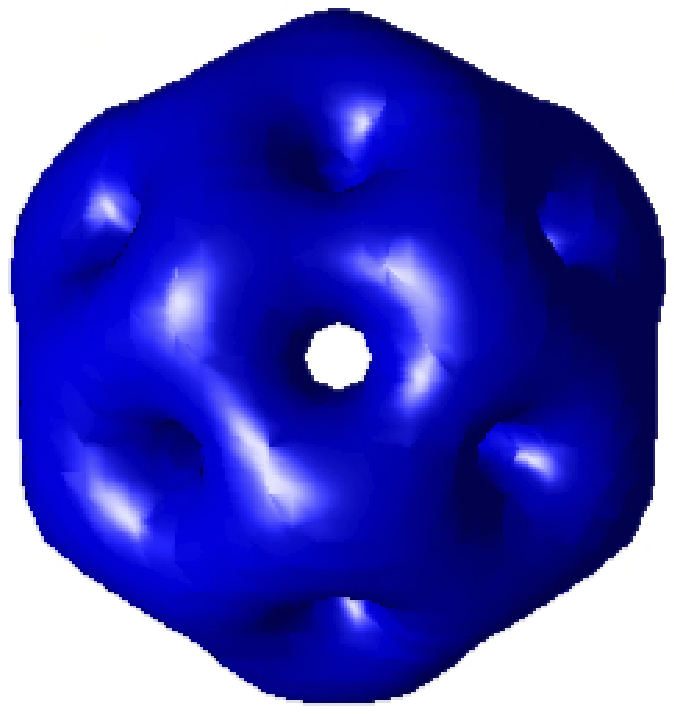}                      
\caption{Baryon density isosurfaces of the rescaled (approximate) Skyrmions 
for $B=6$, $9$, $10$ and $11$ Skyrmions with $m=0$ (first row) and 
$m=1$ (second row). Each corresponds to a value of $\widetilde{\cal B}=0.035$. } 
\label{fig:Bdensities}
\end{figure}

We now describe our procedure for finding these configurations, and 
rescaling them. Initially, a rational map is introduced with a given 
topological charge $B$ and assumed symmetry. Then, the energy is 
minimized (using a simulated annealing process) with respect to the parameters 
appearing in the rational map. In practice, this means minimizing an
integral ${\cal I}$, given in \cite{HMS}, which depends
only on the rational map $R$. Using this familiar approach, we have
re-derived the results obtained in \cite{HMS}, \cite{BSf} and \cite{BMSW}.
The rational maps we obtain are the following: \\

{\large ${\it  B=6}$}\\
The rational map has $D_{4d}$ symmetry, with the $D_4$ generated by
$z\mapsto i z$ and $z\mapsto 1/z$, and has the functional form
\be
R(z)=\fr{z^4+i\al}{z^2\left(i \al z^4+1\right)} \,.
\label{B6}
\ee
The energy is minimized when $\al= 0.15853$. The reflection symmetry 
arises because the parameter $\al$ is real.\\

{\large ${\it  B=9}$}\\
The energy minimizing rational map in this case has $D_{4d}$ symmetry and has 
the form
\be
R(z)=\fr{z\left(\al+i\bt z^4+z^8\right)}{1+i\bt z^4+\al z^8} \,,
\label{B9}
\ee
where $\al=-3.37764$ and $\bt=11.20311$.\\

{\large  ${\it B=10}$ }\\
The minimal energy Skyrmion for $m=1$ is found to have $D_{2h}$
symmetry \cite{BSm2}. This symmetry is consistent with the 
$\al$-particle model, where the field configuration consists 
schematically of a pair of cubic $B=4$ Skyrmions separated by two 
$B=1$ Skyrmions. The optimal rational map with this symmetry is of the form
\begin{equation}
R(z)=\fr{\al+\bt z^2+\gamma z^4+\delta z^6+\epsilon z^8+z^{10}}
{1+\epsilon z^2+\delta z^4+\gamma z^6+\bt z^8+\al z^{10}} \,,
\label{B10}
\end{equation} 
with $\al=0.2772$, $\bt=-9.3594$, $\gamma=14.81$, $\delta=4.977$ 
and $\epsilon=3.015$.\\

{\large ${\it B=11}$}\\
The rational map here is $D_{3h}$-symmetric, and of the form
\begin{equation} 
R(z)=\fr{z^9+\al z^6+\bt z^3+\gamma}
{z^2\left(\gamma z^9+\bt z^6+ \al z^3 +1\right)}
\end{equation}
where $\al=-2.4719$, $\bt=-0.8364$ and $\gamma=-0.1264$.\\

For each of these optimised rational maps, the profile function 
$f(r)$ is obtained by solving the radial equation 
\begin{equation}
f''\left(1+\frac{2B\sin^2 f}{r^2}\right) + \frac{2f'}{r} 
+\frac{\sin2f}{r^2}\left( B\left(f'^2-1\right)
-\frac{{\cal I}\sin^2f}{r^2}\right) - m^2 \sin f = 0\,,
\label{prof}
\end{equation}
using a {\it shooting method}. $B$ is the baryon number and 
${\cal I}$ is the integral mentioned above, evaluated on the
optimised rational map. To avoid singularities at the origin,
the so-called local analysis (see, for details, Reference \cite{Orzag}) 
has been applied. Near $r=0$, the profile function can be approximated 
by the Frobenius series  $f(r)=\pi-a\,r^{\sigma}$
where the indicial exponent $\sigma$ is determined by taking 
the positive root of the quadratic equation $\sigma^{2}+\sigma-2B=0$,
and $a$ is the shooting parameter. The other boundary condition is
$f(\infty)=0$. 

The Skyrme field is then transformed back from Riemann 
to spherical coordinates, i.e. $(r,z,\bar{z})\rightarrow
(r,\theta,\varphi)$, by inverting the equation 
$z=\tan\left(\theta/2\right)\,e^{i\varphi}$, and the current
components $R_i$ are calculated. Numerical integrations, for 
evaluating the contributions to the energies $E$ and $\widetilde{E}$,
and to the scaling integrals ${\cal I}_i$ and $\widetilde{\cal I}_i$, 
are performed by the {\it Gauss-Kronrod} method. 

Finally the values of the rescaling parameters $\lm_i$ 
are obtained by minimizing the energy (\ref{rene}) using 
(once more) a simulated annealing process. They are presented in 
Tables \ref{lambda_without_m} and \ref{lambda_with_m}.
The values are slightly further from 1 for $m=1$ than for $m=0$.
\begin{table}[pht]
	\centering
		\begin{tabular}{ | c | c |  c |  c |c | }
		\hline
		
  $m=0$ &       $B=6$    &     $B=9$    &     $B=10$   & $B=11$ \\
		\hline
$\lm_1$ & $0.98954$  & $1.01272$  & $1.02050$ & $1.00441$  \\ \cline{1-5}
$\lm_2$ & $0.98954$  & $1.01272$  & $0.98811$ & $1.00441$  \\ \cline{1-5}
$\lm_3$ & $1.02118$  & $0.97524$  & $0.99149$ & $0.99103$  \\ \cline{1-5}
\end{tabular}
\caption{Values of the scaling parameters $\lm_i$ for Skyrmions 
with massless pions.} 
\label{lambda_without_m}
\end{table}

\begin{table}[pht]
	\centering
		\begin{tabular}{ | c | c |  c |  c |c | }
		\hline
		
$m=1$ & $B=6$     &     $B=9$       &     $B=10$  &    $B=11$ \\
		\hline
$\lm_1$ & $0.98921$  & $1.01427$   & $1.02224$ & $1.00455$ \\ \cline{1-5}
$\lm_2$ & $0.98921$  & $1.01427$   & $0.98785$ & $1.00455$ \\ \cline{1-5}
$\lm_3$ & $1.02226$  & $0.97268$   & $0.99052$ & $0.99095$ \\ \cline{1-5}
\end{tabular}
\caption{Values of the scaling parameters $\lm_i$ for Skyrmions 
with massive ($m=1$) pions.}
\label{lambda_with_m}
\end{table}

\begin{table}[pht]
\centering
\begin{tabular}{ | c | c |  c |c||c|c| } 		
		\hline
  $B$ & $G$  & $E(m=0)/B$& $\tilde{E}(m=0)/B$&$E(m=1)/B$& $\tilde{E}(m=1)/B $\\
		\hline
$6$   & $D_{4d}$   &  $1.13726$ & $1.13675$ & $1.33151$ & $1.33092$\\ \cline{1-6}
$9$   & $D_{4d}$   &  $1.11681$ & $1.11610$ & $1.31788$ & $1.31696$\\ \cline{1-6}
$10$ & $D_{2h}$   &  $1.11218$ & $1.11172$ & $1.31561$ & $1.31504$\\ \cline{1-6}
$11$ & $D_{3h}$   &  $1.10976$ & $1.10968$ & $1.31632$ & $1.31622$\\ \cline{1-6}
\end{tabular}
\caption{Values of energy per baryon before rescaling (\ref{gene}) and
  after rescaling (\ref{rene}), for $m=0$ and $m=1$. $G$ is the
  symmetry group of the Skyrmion.}
\label{energies}
\end{table}

\begin{table}[pht]
{\footnotesize \centering
\begin{tabular}{ | c | c |  c ||c|c||c|c||c|c| } 		
		\hline
  $\!\!m=0\!\!$ & $\!\!\!\!B=6\!\!\!\!$& $\widetilde{B}=6$ 
& $B=9$& $\widetilde{B}=9$ &$B=10$  & $\widetilde{B}=10$& $B=11$&
$\widetilde{B}=11$\\
		\hline
${\cal I}_1$ & $0.00793$  & $-1.33\,10^{-7}  $ &  $-0.00934$  
&  $1.59\,10^{-9}$ &  $0.00882$ & $1.29\,10^{-9}$    
& $-0.00332$ & $5.11\,10^{-8}$\\ \cline{1-9}
$\!\!\!\!{\cal I}_2\!\!\!\!\!\!$ & $\!\!\!\!0.00793\!\!\!\!$ 
& $-1.05\,  10^{-7}$  & $-0.00934$  &  $ -3.90\, 10^{-12}$ & $-0.01509$   
& $3.74\,10^{-9}$   & $-0.00332$   & $-1.11\,10^{-10}$\\ \cline{1-9}
${\cal I}_3$ & $-0.01593$  & $1.21 \, 10^{-9} $  
& $0.01866$ &  $1.04\,10^{-9}$ &  $0.00624$   
& $8.44\,10^{-10}$ & $0.00662$ & $-3.27\,10^{-11}$\\ \cline{1-9}
\end{tabular}}
\caption{Values of the scaling integrals  (\ref{I1})-(\ref{I3}) before rescaling 
  and after  rescaling, for $m=0$.}
\label{Ident1}
\end{table}

\begin{table}[pht]
{\footnotesize \centering
\begin{tabular}{ | c | c |  c ||c|c||c|c||c|c| } 		
		\hline
  $\!\!m=1\!\!$& $\!\!\!\!\!\!B=6\!\!\!\!\!\!$ 
& $\!\!\!\!\!\!\!\!\!\!\!\!\!\!\widetilde{B}=6\!\!\!\!\!\!\!\!\!\!\!\!\!\!$ 
& $\!\!\!\!B=9\!\!\!\!$& $\widetilde{B}=9$ 
&$\!\!\!\!\!\!\!\!B=10\!\!\!\!\!\!\!\!\!\!$  
& $\!\!\!\!\!\!\widetilde{B}=10\!\!\!\!\!\!$& $B=11$&
$\widetilde{B}=11$\\
		\hline
$\!\!{\cal I}_1\!\!$ & $\!\!0.00882\!\!\!$ 
& $\!\!\!\!\!\!\!\!\!\!\!\!-8.15 \,10^{-7}\!\!\!\!\!\!\!\!\!\!\!\!$  
& $-0.01094$   &  $\!\!\!\!\!\!-6.73\, 10^{-9}\!\!\!\!\!\!$
& $\!\!\!\!\!\!\!\!\!\!0.00975\!\!\!\!\!\!\!\!\!\!$
&$\!\!\!\!\!\!\!\!\!\!-2.61\,10^{-9}\!\!\!\!\!\!\!\!\!\!$
&$-0.00357$&$6.74\,10^{-9}$    \\ \cline{1-9}
${\cal I}_2$ & $0.00882$  & $-7.80 \,10^{-7}$  & $-0.01094$     
&  $-1.45\,10^{-9}$&$-0.01734$& $7.01\,10^{-10}$&$-0.00357$
&$-2.06\,10^{-8}$ \\ \cline{1-9}
${\cal I}_3$ & $-0.01765$  & $-2.27\,10^{-9}$  
& $\!\!\!\!0.02188\!\!\!\!$  &  $-9.31\,10^{-10}$& $0.00759$ 
& $4.16\,10^{-10}$&$0.00715$&$-6.14\,10^{-9}$\\ \cline{1-9}
\end{tabular}}
\caption{Values of the scaling integrals  (\ref{I1})-(\ref{I3}) before rescaling
   and after rescaling , for $m=1$.}
\label{Ident2}
\end{table}

The energies of the approximate Skyrmions, before and after rescaling,
are presented (up to five decimal places) in Table \ref{energies}, 
while the values of the scaling integrals are presented in Tables
\ref{Ident1} and \ref{Ident2} and prove that after rescaling, the 
scaling identities are almost exactly satisfied. This is a check on
our numerics. Note also that $\lm_1\lm_2\lm_3$ is very close to 1, as
anticipated. The changes in the energies are rather small, and much smaller
than the 1\% - 2\% differences in energy between the approximate
Skyrmions before rescaling and the exact Skyrmion solutions. That implies that
rescaling accounts for only a small part of the required change of the
field needed to reach the exact solutions.
  Note also that nuclei with even baryon numbers have larger binding energies than 
those with odd ones. Encouragingly, Skyrmions with even baryon number
generally have lower energy per baryon than Skyrmions with odd 
baryon number. This is seen in our results for $B=9, 10$ and $11$ presented in Table \ref{energies}.

\begin{figure}[pht]
  \centering
   \includegraphics[width=4cm]{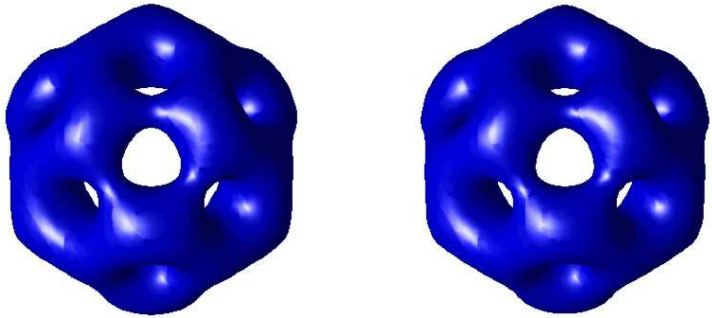}\hs\hs\hs
   \includegraphics[width=4cm]{B=11.eps}                      
\caption{Isosurfaces of the unrescaled (left) and rescaled (right)
baryon density for the $B=11$ Skyrmion with $m=0$.
Each corresponds to a value of $\widetilde{\cal B}=0.035$. }
\label{fig: unBvsreB}
\end{figure}

Finally, baryon density isosurfaces are used to visualize 
the rescaled Skyrmion configurations and these are presented in 
Figure \ref{fig:Bdensities}. The shapes of the 
rescaled configurations are not very different from the unrescaled 
ones, since the values of the scaling parameters are close to $1$. 
This can be seen in Figure \ref{fig: unBvsreB}, where plots of the 
unrescaled and rescaled baryon density isosurfaces of the $B=11$ 
Skyrmion with massless pions are presented.
In order to illustrate more clearly the deformation of the Skyrmions under rescaling we
show in Figure \ref{fig: B=6,9}  the $B=6$  massless Skyrmion 
with the true rescaling parameters changed to the values $\lm_1=\lm_2=0.9$ and $\lm_3=1/\lm_1\lm_2$, and  the $B=9$  massless Skyrmion  with  the rescaling parameters changed to  $\lm_1=\lm_2=1.1$ and $\lm_3=1/\lm_1\lm_2$.

\begin{figure}[pht]
  \centering
   \includegraphics[width=4cm]{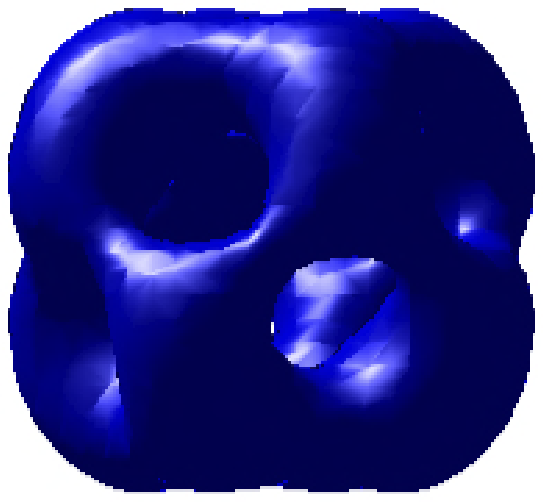}
   \includegraphics[width=4cm]{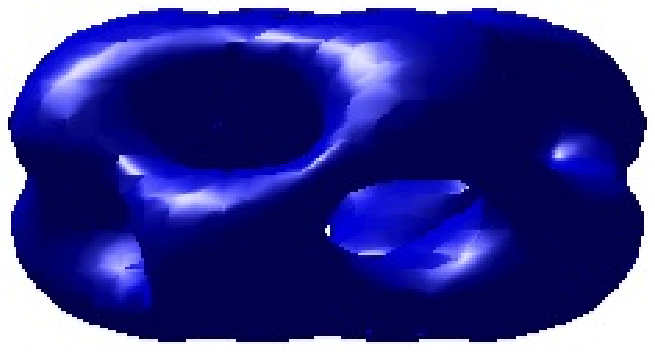}\\
    \includegraphics[width=4cm]{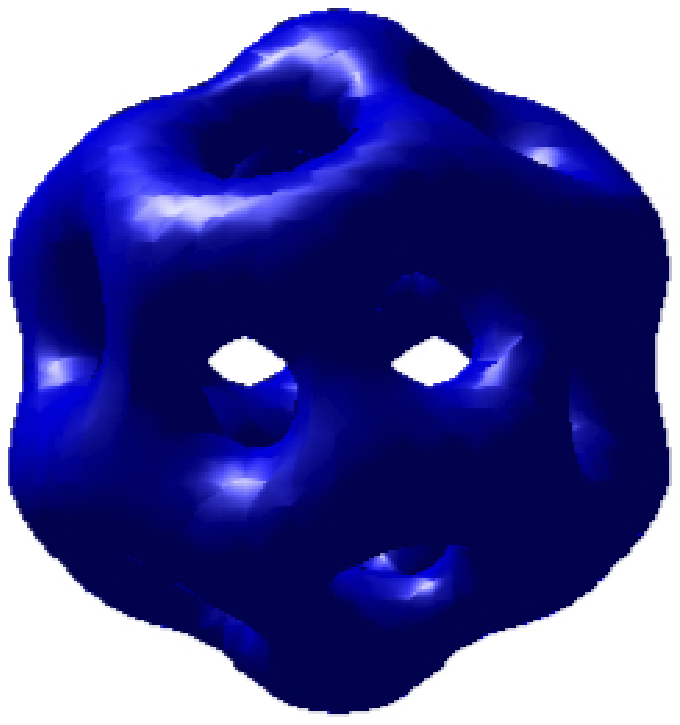}
   \includegraphics[width=4cm]{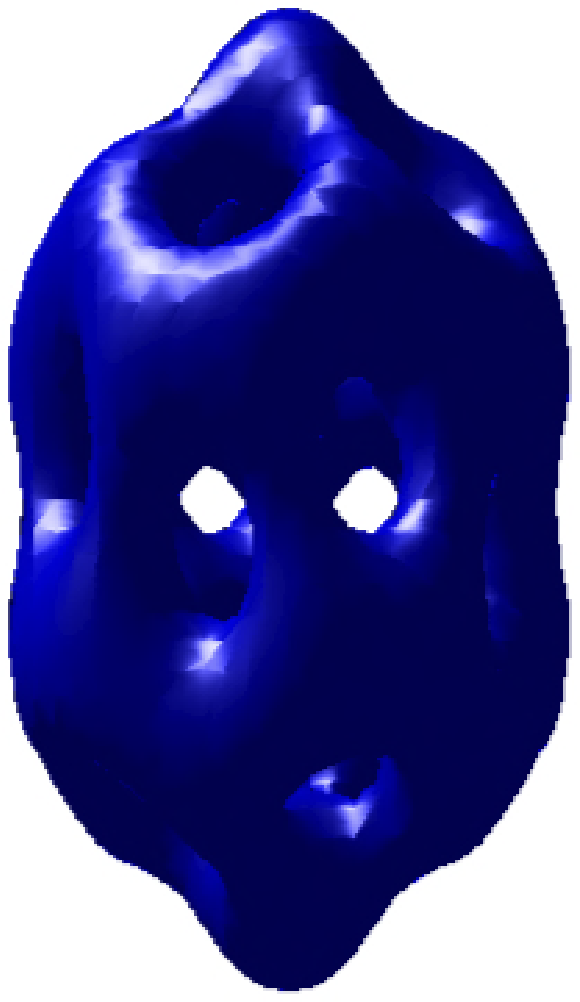}                                           
\caption{Isosurfaces of the {\it true} rescaled (left) and  {\it exaggerated} rescaled (right)
baryon density at a value of the baryon density which is one
quarter of the maximum (i.e. $ 0.25\,\widetilde{\cal B}_{\rm max}$),  for the $B=6$ and the $B=9$ Skyrmion with $m=0$.}
\label{fig: B=6,9}
\end{figure}

\section{Conclusions}

We have started with a Skyrme field configuration of low but not minimal 
energy (\ref{gene}), for selected values of $B$. This is 
constructed using the rational map ansatz. We have then 
rescaled the field, obtaining the energy (\ref{rene}), and have
minimized this numerically, thereby finding a lower energy field 
configuration with the same symmetry. The rescaled field satisfies the
scaling identities (\ref{I1})-(\ref{I3}), or equivalently
(\ref{tildeI1})-(\ref{tildeI3}), and is therefore closer 
to a true Skyrmion solution. In particular, we learn from the values 
of the rescaling parameters whether the true Skyrmion is
more oblate or prolate than the rational map approximation suggests.
This change of shape affects (slightly) the moments of inertia, 
which in turn will affect estimates of the energy spectrum of the quantized 
rotating Skyrmion, modelling real nuclei. 

The $B=6$ configuration shrinks along the $x_3$-axis and stretches in
the $(x_1,x_2)$-plane. It has been observed previously that the
rational map approximation to the $B=6$ Skyrmion is slightly prolate,
and rather too much so, in the sense that its classical electric 
quadrupole moment is positive, and rather too large to match the 
measured quadrupole moment of Lithium-6 \cite{MW}. The rescaling we have found 
makes the approximate Skyrmion less prolate, and this is better for 
fitting the quadrupole moment (although we have not recalculated
it). Unfortunately this conclusion is not totally convincing, because if one
compares the moments of inertia of the approximate $B=6$ Skyrmion
obtained using the rational map ansatz \cite{MW} with the moments of
inertia of the exact solution \cite{BMSW}, then it appears that the
exact solution is more prolate. 

The $B=9$ configuration stretches 
along the $x_3$-axis and shrinks in the $(x_1,x_2)$-plane, so it 
becomes more prolate. Similarly the $B=11$ configuration stretches along 
the $x_3$-axis and shrinks in the $(x_1,x_2)$-plane, becoming prolate 
under rescaling. These are novel insights, and suggest in particular that the 
$B=11$ solution is not closely related in shape to the known $B=12$ 
solution with $D_{3h}$ symmetry \cite{BMSW}, which is significantly
oblate. For $B=10$ the three scaling parameters are independent,
because the field configuration is triaxial, which agrees with the 
structure of the exact $B=10$ Skyrmion (with $m=1$) found numerically.

\clearpage
\vskip 20pt
{\bf Acknowledgements}
\vspace{.25cm}

We thank B. Kleihaus, B. Piette and P. Sutcliffe for useful discussions.  
E.G.C. especially thanks B. Kleihaus for providing the Matlab code
with which the isosurfaces of the baryon density were obtained; he
also thanks K. Kokkotas, S. Massen, N. Vlachos and K. Zagkouris 
for helpful discussions and gratefully acknowledges funding from 
the Division of Mathematics (AUTH).

\end{document}